\documentclass[prd,preprint,aps,showpacs]{revtex4}
\usepackage{graphics}
\newcommand{\newc}{\newcommand}
\newc{\gsim}{\lower.7ex\hbox{$\;\stackrel{\textstyle>}{\sim}\;$}}
\newc{\lsim}{\lower.7ex\hbox{$\;\stackrel{\textstyle<}{\sim}\;$}}
\newc{\gev}{\,{\rm GeV}}
\newc{\mev}{\,{\rm MeV}}
\newc{\ev}{\,{\rm eV}}
\newc{\kev}{\,{\rm keV}}
\newc{\tev}{\,{\rm TeV}}
\newc{\mz}{m_Z}
\newc{\mpl}{M_{Pl}}
\newc{\chifc}{\chi_{{}_{\!F\!C}}}
\newc\order{{\cal O}}
\newc\CO{\order}
\newc\CL{{\cal L}}
\newc\CY{{\cal Y}}
\newc\CH{{\cal H}}
\newc\CM{{\cal M}}
\newc\CF{{\cal F}}
\newc\CD{{\cal D}}
\newc\CN{{\cal N}}
\newc{\eps}{\epsilon}
\newc{\re}{\mbox{Re}\,}
\newc{\im}{\mbox{Im}\,}
\newc{\invpb}{\,\mbox{pb}^{-1}}
\newc{\invfb}{\,\mbox{fb}^{-1}}
\newc{\yddiag}{{\bf D}}
\newc{\yddiagd}{{\bf D^\dagger}}
\newc{\yudiag}{{\bf U}}
\newc{\yudiagd}{{\bf U^\dagger}}
\newc{\yd}{{\bf Y_D}}
\newc{\ydd}{{\bf Y_D^\dagger}}
\newc{\yu}{{\bf Y_U}}
\newc{\yud}{{\bf Y_U^\dagger}}
\newc{\ckm}{{\bf V}}
\newc{\ckmd}{{\bf V^\dagger}}
\newc{\ckmz}{{\bf V^0}}
\newc{\ckmzd}{{\bf V^{0\dagger}}}
\newc{\X}{{\bf X}}
\newc{\bbbar}{B^0-\bar B^0}

\newc{\sgn}{\mbox{sgn}\,}
\newc{\m}{{\bf m}}
\newc{\msusy}{M_{\rm SUSY}}
\newc{\munif}{M_{\rm unif}}
\newc{\slepton}{{\tilde\ell}}
\newc{\Slepton}{{\tilde L}}
\newc{\sneutrino}{{\tilde\nu}}
\newc{\selectron}{{\tilde e}}
\newc{\stau}{{\tilde\tau}}
\def\beq{\begin{equation}}
\def\eeq{\end{equation}}
\def\bea{\begin{eqnarray}}
\def\eea{\end{eqnarray}}
\newc{\ie}{{\it i.e.}}          \newc{\etal}{{\it et al.}}
\newc{\eg}{{\it e.g.}}          \newc{\etc}{{\it etc.}}
\newc{\cf}{{\it c.f.}}
\def\beqn{\begin{eqnarray}}
\def\eeqn{\end{eqnarray}}
\relax
\newcommand{\ba}[1]{\begin{array}{#1}}
\def\ea{\end{arroy}}

\def\beq{\begin{equation}}
\def\eeq{\end{equation}}
\def\bea{\begin{array}}
\def\eea{\end{array}}

\def\to{\rightarrow}

\def\[{\left[}
\def\]{\right]}
\def\({\left(}
\def\){\right)}

\def\U1em{{U(1)_{\rm em}}}
\def\to{\rightarrow}

\def\CL{{\cal C}_L}

\def\sq2{\sqrt{2}}

\def\End{\end{document}}

\def\Dsl{\,\raise.65ex\hbox{/}\mkern-03.5mu D} 
\def\delsl{\raise.15ex\hbox{/}\kern-.57em\partial}
\def\Ksl{\hbox{/\kern-.6700em\rm K}}
\def\Asl{\hbox{/\kern-.1500em \rm A}}
\def\Qsl{\hbox{/\kern-.6000em\rm Q}}
\def\gradsl{\hbox{/\kern-.6501em$\nabla$}}
\def\bar#1{\overline{#1}}

\begin{document}
\draft
\title{Effects of charged Higgs bosons in the deep inelastic
process $\nu_{\tau} {\cal N} \to \tau^- X$ and the possibility of
detecting tau-neutrinos at cosmic neutrino detectors}
\author{M.I. Pedraza-Morales$^{(a)}$, A. Rosado$^{(a)}$ and H. Salazar$^{(b)}$}
\address{$^{(a)}$ Instituto de F\'{\i}sica, BUAP. Apdo. Postal
J-48, C.P. 72570 Puebla, Pue., M\'exico,\\
$^{(b)}$ FCFM-BUAP. Apdo. Postal 1364, C.P. 72000 Puebla, Pue.,
M\'exico}
\date{\today}
\begin{abstract}
We study the deep inelastic process $\nu_{\tau} + {\cal N} \to
\tau^{-} + X$ (with ${\cal N} \equiv (n+p)/2$ an isoscalar nucleon),
in the context of the two Higgs doublet model type II (2HDM(II)). We
discuss the contribution to the total cross section of diagrams, in
which a charged Higgs boson is exchanged. We present results which
show the strong dependence of such contribution on $\tan\beta$ and
$M_{H^{\pm}}$. We show that in the region $50 \leq \tan\beta \leq
200$ and 90 GeV $\leq M_{H^{\pm}}\leq$ 600 GeV with the additional
experimental constraint on the involved model parameters
$M_{H^{\pm}} \geq 1.5 \times \tan\beta$ GeV, the contribution of the
charged Higgs boson exchange diagrams to the cross section of the
charged current inclusive $\nu_{\tau} {\cal N}$ collision can become
important. We obtain that this contribution for an inclusive
dispersion generated through the collision of an ultrahigh energy
tau-neutrino with $E_{\nu} \approx 10^{20}$ eV on a target nucleon
can be larger than the value of the contribution of the $W^{\pm}$
exchange diagrams, provided that $M_{H^{\pm}} \approx 300$ GeV and
$\tan\beta \approx 200$. Such enhancement and the induced variation
on the mean inelasticity $\langle y \rangle^{CC}$ could lead to
sizeable effects in the acceptance of cosmic tau-neutrino detectors
at experiments such as HiRes, PAO, and the CRTNT, which are anchored
to the ground, and at experiments such as EUSO and OWL, which are
proposed to orbit around the Earth. We also compare the contribution
to $\sigma^{tot}_{H^+}$ from the different allowed initial quarks
and we show that the contribution from the bottom quark dominates by
far. This means that the $H^{\pm}$ contribution practically always
gives a top quark in the final state. Such a large component of the
cross section having a top quark event in the final state could have
recognizable features in the EAS experiments.

\end{abstract}
\maketitle

\setcounter{footnote}{0}
\setcounter{page}{2}
\setcounter{section}{0}
\setcounter{subsection}{0}
\setcounter{subsubsection}{0}

\section{Introduction.}

Although the Standard Model (SM) \cite{stanmod}, of the strong and
electroweak interactions describes correctly Particle Physics at
energy ranges currently attainable, one of its basic ingredients,
the scalar Higgs sector, still remains untested. In the SM, the
Higgs sector consists of a single $SU(2)$ doublet, and after
spontaneous symmetry breaking (SSB) it remains a physical state, the
Higgs boson ($h^0_{sm}$), whose mass is not predicted in the theory.
On the other hand, the SM is not expected to be the ultimate
theoretical structure responsible for electroweak symmetry breaking
(EWSB) \cite{HHG,TSM}. One of the most simple extensions of the SM
is the so called two Higgs doublet model (2HDM). There are four
classes of 2HDM which naturally avoid tree-level flavor-changing
neutral currents that can be induced by Higgs boson exchange
\cite{Barger:1989fj}. These models include a Higgs sector with two
scalar doublets, which give masses to the up and down-type fermions
as well as the gauge bosons. Model II is particularly interesting,
where one of the Higgs scalar doublet couples to the up-components
of isodoublets while the other to the down-components. Model II is
that utilized in SUSY theories. The 2HDM(II) has a rich structure
and predicts interesting phenomenology \cite{HHG}. The physical
spectrum consists of two neutral CP-even states ($h^{0},H^{0}$) and
one CP-odd ($A^{0} $), as well as a pair of charged scalar particles
($H^{\pm }$). The advantage of such model is the fact that any Higgs
sector built only upon doublets preserves naturally the lowest-order
electroweak relation $\rho=1$, with $\rho=M^2_{W^{\pm}}/(M^2_Z
\cos^2\theta)$, which has been tested with a good accuracy. On the
phenomenological side, an important aspect of the 2HDM is that the
Higgs sector may provide an additional source of CP violation
\cite{cpv}.

Several experimental lower limits on the charged Higgs boson mass
in this model have been reported in the literature:
\begin{equation}\label{bound1}
M_{H^{\pm}} > 79.3 \, \mbox{GeV \hspace{1.0cm} (95\% C.L.) \cite{Heister:2002ev}} \, ,\\
\end{equation}
\begin{equation}\label{bound2a}
M_{H^{\pm}} > (0.97,\,1.28,\,1.89) \times \tan\beta \, \mbox{GeV
\hspace{1.0cm} (95\% C.L.) \cite{Ackerstaff:1998yk},
\cite{Abbiendi:2002jw}, \cite{Abbiendi:2001fi}} \, , \\
\end{equation}
\begin{equation}\label{bound2b}
M_{H^{\pm}} > (0.97,\,1.5,\,1.9,\,2.5,\,2.6) \times \tan\beta \,
\mbox{GeV \hspace{0.5cm} (90\% C.L.) \cite{Ammar:1996xh},
\cite{Stahl:1996gu}, \cite{Buskulic:1994gj}, \cite{Barate:2000rc},
\cite{Acciarri:1996bv}} \, . \\
\end{equation}

\noindent Further, based on the discussions on $\tan\beta$ given in
Refs. \cite{Roy:2005yu} and \cite{Baer:2002hf}, we restrict
ourselves by taking the following upper limit on $\tan\beta$
\begin{equation}\label{bound3}
\tan\beta \leq 200 \, .\\
\end{equation}

Large-scale neutrino telescopes \cite{telescopes} have as a main
goal the detection of ultrahigh-energy (UHE) cosmic neutrinos
($E_\nu \geq 10^{12}$ eV) produced outside the atmosphere
(neutrinos produced by galactic cosmic rays interacting with
interstellar gas, and extragalactic neutrinos)
\cite{uhe-neut,Gandhi:1998ri}. UHE neutrinos can be detected by
observing long-range muons and tau-leptons decays produced in
charged-current neutrino-nucleon interactions. UHE tau-neutrinos
are generated through neutrino oscillations
\cite{Blanch1,Bertou:2001vm}. The detection of UHE neutrinos will
provide us with the possibility to observe $\nu {\cal
N}$-collisions with a neutrino energy in the range $10^{12}$ eV
$\leq E_\nu \leq 10^{21}$ eV and a target nucleon at rest. An
enlightening discussion on UHE neutrino interactions is given by
R. Gandhi {\it et al.} \cite{Gandhi:1998ri}.

We discuss in this paper the cross section of the deep inelastic
process $\nu_{\tau} + {\cal N} \to \tau^{-} + X$ (${\cal N} \equiv
(n+p)/2$ an isoscalar nucleon), in the context of the SM and of the
2HDM(II). We perform our numerical calculations using the parton
model \cite{partmod1,partmod2} with the parton distribution
functions reported by J. Pumplin {\it et al.} \cite{pumplin}. We use
the CTEQ PDFs provided in an $n_f=5$ active flavors scheme. Our aim
is to calculate how large can be the contribution of diagrams, in
which a charged Higgs boson is exchanged, to the total cross section
of the mentioned inclusive process in the frame of the 2HDM(II). In
the 2HDM(II) the couplings of the down-type quarks and charged
leptons are proportional to $m_{f} \times \tan\beta$. Hence, for
large $\tan\beta$ the contribution of $H^{\pm}$-exchange diagrams
will be maximal in this model.

This paper is organized as follows. In section II, we review the
notation and physical region of the inclusive $\nu_{l} {\cal
N}$-dispersion. In section III, we present the formulae for the
differential cross section of the deep inelastic process $\nu_{\tau}
{\cal N} \to \tau^- + X$, in the context of the SM and the 2HDM(II).
In section IV, we give and discuss our results for the total cross
section rates of the charged current deep inelastic process
$\nu_{\tau} \, {\cal N}$ in the frame of the SM and 2HDM(II). In
section V, we discuss the contribution of charged Higgs boson to the
charged current cross section of the $\nu_{\tau} {\cal N}$
collision, the induced variation on the mean inelasticity $\langle y
\rangle^{CC}$ and hence the effects on the possibility of detecting
cosmic tau-neutrinos. Finally, in section VI, we summarize our
conclusions.

\section{Notation, Physical Region}

\par In this section we review the notation and the physical
region for the parameters of the inclusive process
\begin{equation}\label{proc1}
\nu_{l} + {\cal N} \to  l^{-} + X \, ,
\end{equation}
\noindent where $\nu_{l}, \, {\cal N}$, and $l^{-}$ stand for the
incoming neutrino, the target nucleon and the outgoing lepton
$l^-$, respectively. We will denote the four-momenta of these
particles by $p$, $P_{\cal N}$ and $p'$, respectively. In
accordance with the kinematics for the collision of a neutrino on
a target nucleon, the following construction is chosen:
\begin{equation}\label{momdef}
p^\mu=E_{\nu}(1,0,0,1) \, , \hspace{2.5cm} P^\mu_{\cal N}=M_{\cal N}(1,0,0,0) \, .\\
\end{equation}

\par As usual, we define the invariants \cite{partmod2}:
\begin{equation}\label{invar}
\begin{array}{lll}
s&=&(p+P_{\cal N})^2 \, , \\
Q^2&=&-(p-p')^2 \, , \\
\nu&=&P_{\cal N}(p-p')/M_{\cal N} \, , \\
\end{array}
\end{equation}
\noindent and the dimensionless variables:
\begin{equation}\label{dimless}
x=\displaystyle\frac{Q^2}{2 \nu M_{\cal N}} \, , \hspace{1cm}
y=\displaystyle\frac{\nu}{E_{\nu}} \, . \hspace{1cm}
\end{equation}
\noindent The physical region of these kinematical variables is
obtained by requiring that the scalar products of any two particle
four-momenta be positive and the determinant $\Delta_3$ of the three
independent four-momenta (whenever possible we will neglect the
fermion masses):
\begin{equation}\label{del3}
\nonumber\\[1mm]
\Delta_3(p,p',P_{\cal N})= \left|
\begin{array}{ccc}
0 & pp' & pP_{\cal N}\\
p'p & 0 & p'P_{\cal N}\\
P_{\cal N}p & P_{\cal N}p' & 0\\
\end{array}
\right|
\nonumber\\[1mm]
\end{equation}
\noindent be positive \cite{Kinematics}.

\par From the non-negative character of the scalar products we  find:
\begin{equation}\label{physreg1}
0 \leq x  \leq 1 \, ,\hspace{0.75in} 0 \leq y \leq 1 \, .
\end{equation}
Explicit evaluation of $\Delta_3$ using (\ref{invar}),
(\ref{dimless}) and (\ref{del3}) gives:
\begin{equation}\label{expdel3}
\Delta_3 = (s/2)^3 \, 2 x y (1 - y) \, .
\end{equation}
\noindent with $s=2 M_{\cal N} E_{\nu}$. The condition $\Delta_3
\geq 0$ does not lead to additional restrictions on the physical
region.

\noindent The expressions given in (\ref{physreg1}), define the
physical region for the dimensionless variables $x$, and $y$. We
have taken all fermion masses to be zero, which implies that in the
calculation of the total cross section the integration over the
momentum transfer square extends up to zero. However, the parton
distributions can be used only when $Q^2$ is not too small.
Furthermore, in order to separate deep inelastic from elastic
scattering, a cut on the invariant mass $W$ of the unobserved
particles in the final state is required. Therefore, besides the
kinematical conditions (\ref{physreg1}), we also have in general the
following constraints:
\begin{eqnarray}\label{qc2wc}
Q^2&=&sxy\geq Q^2_{c} \, , \nonumber\\
W&=&sy(1-x) \geq W_{c} \, .
\end{eqnarray}
\noindent The cuts for $Q^2$ and $W$ constrain further the
physically allowed region for the process (\ref{proc1}). The
physical region can now be written in terms of the dimensionless
variables as follows:
\begin{eqnarray}\label{phyreg2}
\frac{Q^2_{c}}{s} \leq x \leq 1-\frac{W_{c}}{s} \, , \hspace{0.75in}
\max \left\{\frac{Q^2_c}{sx},\frac{W_{c}}{s(1-x)} \right \} \leq y
\leq 1 \, .
\end{eqnarray}

\noindent The physically allowed region can also be expressed as
\begin{eqnarray}\label{phyreg3}
\frac{Q^2_{c}+W_{c}}{s} \leq y \leq 1 \, , \hspace{0.75in}
\frac{Q^2_c}{sy} \leq x \leq 1 - \frac{W_c}{sy} \, .
\end{eqnarray}

\section{The cross section for the inclusive process $\nu_{\tau} + {\cal N} \to
\tau^- + X$}

\subsection{The differential cross section for the process
$\nu_{\tau} + {\cal N} \to \tau^- + X$ in the SM}

\begin{figure}[floatfix]
\begin{center}
\includegraphics{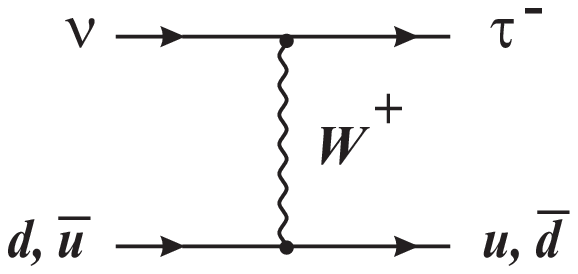}
\caption{Diagrams which at the quark level contribute to the process
$\nu_{\tau} + {\cal N} \to \tau^- + X$ at the lowest order in
$\alpha$ in the SM ($d$ stands for $d$-, $s$- and $b$-quark; and $u$
stands for $u$- and $c$-quark).} \label{fig:figsm}
\end{center}
\end{figure}

The differential cross section for the inclusive reaction
\begin{equation}\label{procsm}
\nu_{\tau}(p) + {\cal N}(P_{\cal N}) \to  \tau^-(p') + X \, ,
\end{equation}
where ${\cal N} \equiv (n+p)/2$ is an isoscalar nucleon, at the
lowest order in $\alpha$ in the frame of the SM (see
Fig.~\ref{fig:figsm}) is given as follows \cite{dis-sm}:
\begin{equation}\label{dcssm}
\frac{d^2\sigma_{sm}}{dxdy}= \frac{2 G^2_F M_{\cal N} E_{\nu}}{\pi}
\left( \frac {M^2_{W^{\pm}}}{Q^2+M^2_{W^{\pm}}} \right)^2 [x
q_W(x,Q^2) + x \bar{q}_W(x,Q^2) (1-y)^2] \, ,
\end{equation}
where $Q^2$, $x$ and $y$ are defined in (\ref{invar}) and
(\ref{dimless}) and $M_{\cal N}$ stands for the nucleon mass. The
quantities $q_W(x,Q^2)$ and $\bar{q}_W(x,Q^2)$ are given as
\begin{eqnarray}\label{fqw}
q_W(x,Q^2) &=& \frac{u_v(x,Q^2) + d_v(x,Q^2)}{2} \nonumber\\
& &+\frac{u_s(x,Q^2) + d_s(x,Q^2)}{2} + s_s(x,Q^2) + b_s(x,Q^2)
\, , \nonumber\\
\bar{q}_W(x,Q^2) &=& \frac{u_s(x,Q^2) + d_s(x,Q^2)}{2} +
c_s(x,Q^2) \, ,
\end{eqnarray}
where the valence and sea parton distribution functions (PDFs),
$q_v(x,Q^2)$ and $q_s(x,Q^2)$, can be expressed as
\begin{eqnarray}\label{fqprot}
u_v(x,Q^2) &=& u(x,Q^2) - \bar{u}(x,Q^2) \, , \nonumber\\
d_v(x,Q^2) &=& d(x,Q^2) - \bar{d}(x,Q^2) \, , \nonumber\\
u_s(x,Q^2)&=&\bar{u}(x,Q^2) \, , \nonumber\\
d_s(x,Q^2)&=&\bar{d}(x,Q^2) \, , \nonumber\\
c_s(x,Q^2) &=& c(x,Q^2)=\bar{c}(x,Q^2) \, , \nonumber\\
s_s(x,Q^2) &=& s(x,Q^2)=\bar{s}(x,Q^2) \, , \nonumber\\
b_s(x,Q^2) &=& b(x,Q^2)=\bar{b}(x,Q^2) \, ,
\end{eqnarray}
where the PDFs $q(x,Q^2)$ describe the quark $q$ content of the
proton. In other words, the parton distribution functions
$q(x,Q^2)$ give the probabilities to find a quark $q$ inside a
proton with the fraction $x$ of the proton momentum: $q^\mu= x
P^\mu$, in a scattering process with momentum transfer square
$Q^2$.

In the case of the standard model the couplings of the fermions to
the $W^{\pm}$ boson are given by the lagrangian
\begin{equation}\label{lagwsm} {\cal L}=-\frac{g}{\sqrt{2}}
\sum_{(f_u,f_d)}\left\{ \left( \bar{f}_u \gamma^{\mu}
\frac{1-\gamma_5}{2} f_d \right) W^+_{\mu} + \left( \bar{f}_d
\gamma^{\mu} \frac{1-\gamma_5}{2} f_u \right) W^-_{\mu} \right\}
\, ,
\end{equation}
where $f_u$ and $f_d$ stand for the up- and down-components of the
fermion doublet.

\subsection{The differential cross section for the process
$\nu_{\tau} + {\cal N} \to \tau^- + X$ in the 2HDM(II)}

\begin{figure}[floatfix]
\begin{center}
\includegraphics{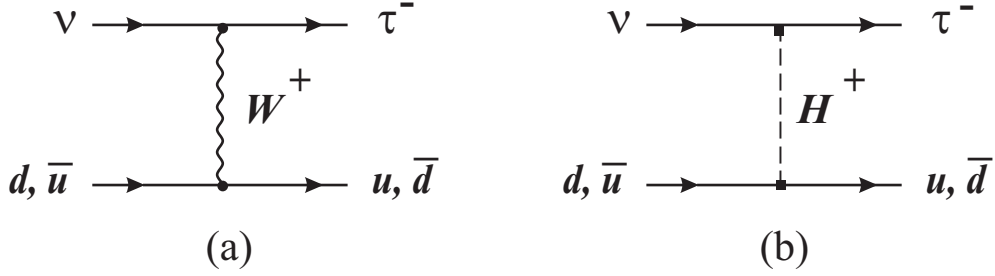}

\caption{Diagrams which at the quark level contribute to the process
$\nu_{\tau} + {\cal N} \to \tau^- + X$ at the lowest order in
$\alpha$ in the 2HDM(II) ($d$ stands for $d$-, $s$- and $b$-quark;
and $u$ stands for $u$- and $c$-quark).} \label{fig:fig2hdm}
\end{center}
\end{figure}

The differential cross section for the inclusive reaction
(\ref{procsm}), at the lowest order in $\alpha$ in the frame of the
2HDM(II) (see Fig.~\ref{fig:fig2hdm}), can be written as follows
\cite{Rosado:2006hn}:
\begin{equation}\label{dcs2hdm1}
\frac{d^2\sigma_{2hdm}}{dxdy}=\frac{d^2\sigma_{sm}}{dxdy} +
\frac{d^2\sigma_{H^+}}{dxdy} \, ,
\end{equation}
where for large $\tan\beta$
\begin{equation}\label{dcs2hdmh}
\frac{d^2\sigma_{H^+}}{dxdy}= \frac{G^2_F M_{\cal N} E_{\nu}}{2 \pi}
\frac {m^2_{\tau} M^2_{W^{\pm}} \tan^4\beta}{(Q^2+M^2_{H^{\pm}})^2}
y^2 [x q_H(x,Q^2) + x \bar{q}_H(x,Q^2)] \, ,
\end{equation}
where $Q^2$, $x$ and $y$ are defined in (\ref{invar}) and
(\ref{dimless}) and $M_{\cal N}$ stands for the nucleon mass. The
quantities $q_W(x,Q^2)$ and $\bar{q}_W(x,Q^2)$ are given in
(\ref{fqw}), whereas $q_H(x,Q^2)$ and $\bar{q}_H(x,Q^2)$ are given
as
\begin{eqnarray}
q_H (x,Q^2) &=& \frac{m^2_d}{M^2_{W^{\pm}}} \left(
\frac{u_v(x,Q^2) + d_v(x,Q^2)}{2} +
\frac{u_s(x,Q^2) + d_s(x,Q^2)}{2} \right) \nonumber\\
&& + \frac{m^2_s}{M^2_{W^{\pm}}} \, s_s(x,Q^2) +
\frac{m^2_b}{M^2_{W^{\pm}}} \,
b_s(x,Q^2) \nonumber\\
\bar{q}_H (x,Q^2) &=& \frac{m^2_d}{M^2_{W^{\pm}}} \left(
\frac{u_s(x,Q^2) + d_s(x,Q^2)}{2} \right) +
\frac{m^2_s}{M^2_{W^{\pm}}} \, c_s(x,Q^2) \, .
\end{eqnarray}

In the case of the 2HDM(II) the couplings of the fermions to the
$W^{\pm}$ boson are given by the lagrangian in Eq.(\ref{lagwsm}), in
a way similar to the SM \cite{HHG}. On the other side, taking the
elements of the CKM-matrix $V_{ij}= \delta_{ij}$, the couplings of
the fermions to the $H^{\pm}$ boson are given by the lagrangian
\cite{HHG}
\begin{equation}\label{lagh2hdm}
{\cal L}=\frac{g}{ M_{W^{\pm}}} \left\{ m_{\tau} \tan\beta \left(
\bar{\nu} \frac{1+\gamma_5}{2} \tau \right) + m_{u} \cot\beta
\left( \bar{u} \frac{1-\gamma_5}{2} d \right) + m_{d} \tan\beta
\left( \bar{u} \frac{1+\gamma_5}{2}d \right) \right\}H^+ + h.c.
\end{equation}

\section{Results for deep inelastic $\nu_{\tau} {\cal N}$ in the SM and
the 2HDM(II)}

We present results for the case of unpolarized deep inelastic
process $\nu_{\tau} + {\cal N} \to \tau^- + X$ with a neutrino
energy in the range $10^{14}$ eV  $\leq E_{\nu} \leq 10^{20}$ eV and
the nucleon at rest, $\it{i.e.}$ a target nucleon. We take $10^{14}$
eV $\leq E_{\nu}$ which leads to a condition where all fermion
masses become negligible with respect to the total energy $s=2
M_{\cal N} E_{\nu}$, even the top quark mass. We take cuts of $\sim
2$ GeV$^2$ and 10 GeV$^2$ for $Q^2$ and the invariant mass $W$,
respectively. These values for the cuts are suited for the parton
distribution functions of J. Pumplin {\it et al.} \cite{pumplin}
which we will use in our calculations. We have checked numerically
that the total cross section rates do not depend on the choice of
the cuts on the momentum transfer square $Q^2$, when they take on
values of a few GeV$^2$. This is due to the fact that the
propagators involved in the calculation of the cross section are
$1/(M^2_{W^{\pm}}+Q^2)$ and $1/(M^2_{H^{\pm}}+Q^2)$

We perform our numerical calculations taking for the quark masses:
$m_u=4$ MeV, $m_d=8$ MeV, $m_c=1.5$ GeV, $m_s=150$ MeV, $m_b=4.9$
GeV and $m_t=174$ GeV. For the evaluation of the $H^+ \tau^-
\nu_{\tau}$ coupling we take $m_{\nu_{\tau}}=0$ and $m_{\tau}=1,777$
MeV. Also, we take $M_{W^{\pm}}=80.4$ GeV for the mass of the
charged boson $W^{\pm}$ \cite{partdata}


Taking into account the constraints on $M_{H^{\pm}}/\tan\beta$ given
by (\ref{bound2a}) and (\ref{bound2b}), we will present results for
three different cases: (I) $M_{H^{\pm}}=1.5 \, \tan\beta$ GeV; (II)
$M_{H^{\pm}}=2.0 \, \tan\beta$ GeV; (III) $M_{H^{\pm}}=3.0 \,
\tan\beta$ GeV. In all these cases, the conditions (\ref{bound1})
and (\ref{bound3}) will be fulfilled. Hence, we restrict our
numerical analysis to the region $50 \leq \tan\beta \leq 200$ and 90
GeV$\leq M_{H^{\pm}} \leq 600$ GeV.

\subsection{Total cross section $\sigma^{tot}_{sm}$ and
$\sigma^{tot}_{2hdm}$}

We display in our numerical results for the total cross section as a
function of $E_{\nu}$ in the range $10^{14}$ eV $\leq E_{\nu} \leq
10^{20}$ eV, with $E_{\cal N}=M_{\cal N}$ in
Tables~\ref{tab:t1}-\ref{tab:t3}. In the second column of these
Tables we give results for the SM. The numerical results for the
2HDM(II) are given in the third-sixth columns of Table~\ref{tab:t1}
(Case I), Table~\ref{tab:t2} (Case II), and Table~\ref{tab:t3} (Case
III).

Tables \ref{tab:t1}, \ref{tab:t2} and \ref{tab:t3} show that the
total cross section $\sigma^{tot}_{2hdm}$ becomes larger as
$\tan\beta$ increases and/or as $M_{H^{\pm}}$ decreases.

\subsection{The ratio $\sigma^{tot}_{H^+}(\nu_{\tau} + {\cal N} \to
\tau^{-} + X)/\sigma^{tot}_{sm}(\nu_{\tau} + {\cal N} \to \tau^{-}
+ X)$}

We present in our results for the ratio
$\sigma^{tot}_{H^+}(\nu_{\tau} + {\cal N} \to \tau^{-} +
X)/\sigma^{tot}_{sm}(\nu_{\tau} + {\cal N} \to \tau^{-} + X)$ as a
function of $E_{\nu}$ in the range $10^{14}$ eV $\leq E_{\nu} \leq
10^{20}$ eV, with $E_{\cal N}=M_{\cal N}$, in Table~\ref{tab:t4}
(Case I), Table~\ref{tab:t5} (Case II), and Table~\ref{tab:t6} (Case
III).

We observe in Tables \ref{tab:t4}, \ref{tab:t5} and \ref{tab:t6}
that the ratio $\sigma^{tot}_{H^+}/\sigma^{tot}_{sm}$ becomes larger
as $\tan\beta$ increases and/or as $M_{H^{\pm}}$ decreases.
Furthermore, we notice that $\sigma^{tot}_{H^+}/\sigma^{tot}_{sm}$
depends on $E_{\nu}$. We see that for $\tan\beta=150$ and
$\tan\beta=200$, the greater $E_{\nu}$, the larger ratio
$\sigma^{tot}_{H^+}/\sigma^{tot}_{sm}$ becomes.

\subsection{Comparison of the contribution to $\sigma^{tot}_{H^+}$ from the different quarks }

Finally, in Table~\ref{tab:t7} we compare the contribution to
$\sigma^{tot}_{H^+}$ from the different allowed initial quarks (see
Fig.~\ref{fig:fig2hdm}(b)), taking $M_{H^{\pm}}=400$ GeV and
$\tan\beta=200$. We observe in this table that the contribution from
the bottom quark dominates by far (a similar behavior is obtained in
the other cases). This fact implies that the contribution of the
$H^{\pm}$ exchange diagrams to the total cross section of the
$\nu_{\tau}\,{\cal N}$ scattering in the frame of the 2HDM(II) is
the same regardless of whether the nucleon is a proton, a neutron or
an isoscalar nucleon, because these particles have the same content
of b-quark. We woul like to emphasize that the $H^{\pm}$
contribution practically always gives a top quark in the final
state, resulting in a much larger values of the mass invariant $W$
and dimensionless parameter $y$ in the 2HDM with respect to the
corresponding values in the case of the SM. Such a large component
of the cross section having a top quark event in the final state
could have other recognizable features in the EAS experiments, which
may be worth looking into.


\section{Charged Higgs boson effects in the possibility of detecting ultra-high
energy tau-neutrinos}

Cosmic neutrino fluxes can initiate air-showers through interaction
in the atmosphere, or in the Earth. Neutrino trajectories will be
going down in a nearly horizontal way in the former case, whereas in
a Earth-skimming manner in the latter case. Thus, it is important to
know the acceptance (event rate/flux) of proposed air shower
experiments for detecting both types of neutrino -initiated events
[Horizontal air-showers (HAS) and Upgoing (Earth-skimming)
air-showers (UAS)]. These acceptances for fluorescence detectors
have been calculated in \cite{Palomares-Ruiz:2005xw} for experiments
as High Resolution Fly's Eye Detector (HiRes) \cite{HIRES}, Pierre
Auger Observatory (PAO) \cite{Augerfluo}, and the Cosmic Ray Tau
Neutrino Telescopes (CRTNT) \cite{TAfluo}, which are anchored to the
ground, and at experiments as Extreme Universe Space Observatory
(EUSO) \cite{EUSO} and Orbiting wide angle light-collectors (OWL)
\cite{OWL}, which are proposed to orbit around the Earth. In
Ref.\cite{Palomares-Ruiz:2005xw}, it is stated that the very
different dependence on the cross section of the HAS (linear) and
UAS (non linear) acceptances will provide a practical method to
measure the charged current cross section of the neutrino nucleon
scattering $\sigma^{CC}_{\nu {\cal N}}$. One simply has to exploit
the ratio of UAS to HAS event rates. However, one assumption made in
Ref.\cite{Palomares-Ruiz:2005xw} was taking that
$\sigma^{CC}_{\nu_{e} {\cal N}}=\sigma^{CC}_{\nu_{\mu} {\cal
N}}=\sigma^{CC}_{\nu_{\tau} {\cal N}}$. This is valid in the SM, but
it is no longer valid in the 2HDM(II), because the
$\sigma^{CC}_{\nu_{\tau} {\cal N}}$ could be twice larger than
$\sigma^{CC}_{\nu_{e} {\cal N}}=\sigma^{CC}_{\nu_{\mu} {\cal N}}$ in
this model. Hence, the neutrino charged current interaction
mean-free path (introduced in Ref.\cite{Palomares-Ruiz:2005xw})
could be for $\nu_{\tau}$ one half of that for $\nu_e$ and
$\nu_{\mu}$, in the mentioned model. Therefore, the existence of
charged Higgs bosons could imply a large deviation from the results
for the acceptances for space based (or ground based) tau-neutrino
detectors presented in Fig. 7 of Ref.\cite{Palomares-Ruiz:2005xw}.

Further, it was already pointed out in Ref.\cite{Zas:2005zz}, that
besides the obvious role of the neutrino cross section in the actual
interaction that leads to the possible neutrino detection there are
other more subtle effects. The $y$ distribution of the cross section
also has an impact on the detection rate
\cite{Zas:2005zz,Gandhi:1995tf,Arteaga:1}. It is clear from
Eq.(\ref{dcssm}) and Eq.(\ref{dcs2hdmh}) that the $y$ distribution
of the charged current interaction could change due to the effects
of charged Higgs bosons. In fact, in Table~\ref{tab:t8} we show how
the mean inelasticity $\langle y \rangle^{CC}$ would change in the
THDM(II) (We present only numerical results for the Case I, in which
the effects are maximal). Our results obtained for $\langle y
\rangle^{CC}$ in the frame of the SM are in good agreement with
those reported in Ref.\cite{Aramo:2004pr}. We observe in
Table~\ref{tab:t8} that $\langle y \rangle^{CC}_{2hdm} \cong 2.5
\times \langle y \rangle^{CC}_{sm}$ for $M_{H^{\pm}}=300$ GeV,
$\tan\beta=200$ and $E_{\nu}= 10^{20}$ eV. Such a drastic variation
in $\langle y \rangle^{CC}$ could induce a large variation in the
possibility of detecting ultra-high energy tau-neutrinos through
inclined showers \cite{Zas:2005zz}. We end this section saying that
the possible effects of charged Higgs bosons on the detection of
cosmic tau-neutrinos deserve future detailed investigations.

\section{Conclusions}

We have calculated the total cross section rates for the deep
inelastic process $\nu_{\tau} + {\cal N} \to \tau^- + X$, where
${\cal N} \equiv (n+p)/2$ is an isoscalar nucleon, in the frame of
the SM and the 2HDM(II). In the case of the 2HDM(II) we have taken
into account the contribution of the diagrams in which a charged
Higgs boson is exchanged $\sigma^{tot}_{H^+}$.

We obtained our numerical results considering the nucleon at rest
and according to the following conditions: taking $M_{W^{\pm}} =
80.4$ GeV, $\sin^2 \theta _W=0.223$, a neutrino energy in the range
$10^{14}$ eV $\leq E_{\nu} \leq 10^{20}$ eV and setting cuts of
$\sim 2$ GeV$^2$ and 10 GeV$^2$ for the momentum transfer square
$Q^2$ and the invariant mass ($W$), respectively. We made use of the
parton distribution functions of J. Pumplin {\it et al.}, those
provided in an $n_f=5$ active flavors scheme. In order to take into
account the experimental data, we restricted our numerical analysis
to the region $50 \leq \tan\beta \leq 200$ and 90 GeV$\leq
M_{H^{\pm}} \leq 600$ GeV, with the additional constraint
$M_{H^{\pm}} \geq 1.5 \times \tan\beta$ GeV.

We have shown that the most important contribution to
$\sigma^{tot}_{H^+}$ comes from the $H^{\pm}$-exchange diagram with
an initial b-quark (and hence an outgoing t-quark). This fact
implies that the contribution of the $H^{\pm}$ exchange diagrams to
the total cross section of the $\nu_{\tau}\,{\cal N}$ scattering in
the frame of the 2HDM(II) is the same regardless of whether the
nucleon is a proton, a neutron or an isoscalar nucleon, because
these particles have the same content of b-quark. On the other hand,
we would like to stress that the $H^{\pm}$ contribution leads to a
top quark in the final state and that such a large component of the
cross section having a top quark event in the final state could have
characteristic aspects in the EAS experiments.

We found that the ratio $\sigma^{tot}_{H^+}/\sigma^{tot}_{sm}$
becomes larger as $\tan\beta$ increases and/or as $M_{H^{\pm}}$
decreases. We also found that the ratio
$\sigma^{tot}_{H^+}/\sigma^{tot}_{sm}$ depends on $E_{\nu}$. In
particular, we obtain that for $\tan\beta=150$ and $\tan\beta=200$:
the greater $E_{\nu}$, the larger ratio
$\sigma^{tot}_{H^+}/\sigma^{tot}_{sm}$ becomes.

We showed that the contribution of the charged Higgs boson exchange
diagrams can lead to a sizeable enhancement with respect to the SM
cross section rates for the charged current $\nu_{\tau} {\cal N}$
deep inelastic scattering. For the case of an ultrahigh energy
tau-neutrino with $E_{\nu}=10^{20}$ eV colliding on a target nucleon
such enhancement reached $110\%$ provided that $\tan\beta=200$ and
$M_{H^{\pm}}=300$ GeV. Besides, we obtain $\langle y
\rangle^{CC}_{2hdm} \approx 2.5 \times \langle y \rangle^{CC}_{sm}$
for the same values of $E_{\nu}$, $\tan\beta$ and $M_{H^{\pm}}$.
This enhancement and the induced variation in the mean inelasticity
$\langle y \rangle^{CC}$ could lead to sizeable effects in the
acceptance of the cosmic tau-neutrino detectors at space based
experiments such as the EUSO and OWL proposals and at ground-based
experiments such as PAO, HiRes and the CRTNT.

\bigskip

\begin{center}
{\bf ACKNOWLEDGMENTS}
\end{center}

The authors are grateful to {\it Sistema Nacional de Investigadores
} and {\it CONACyT} (M\'{e}xico) for financial support and also to
Dr. Antonio Flores-Riveros for a careful and critical reading of the
manuscript. A.R. would like to thank Prof. D.P. Roy for valuable
comments.

\newpage

\begin{table}
\begin{tabular}{|c|c|c|c|c|c|}
\hline
& & \multicolumn{4}{c|}{$\sigma^{tot}_{2hdm}$(cm$^2$)}\\
\cline{3-6} {$E_{\nu}$(eV)}& $\sigma^{tot}_{sm}$(cm$^2$) & I(a)
& I(b) & I(c) & I(d) \\
\hline \hline

$10^{14}$&$\;2.01 \times 10^{-34}\;$&$\;2.17 \times 10^{-34}\;$&$\;2.23 \times 10^{-34}\;$&$\;2.27 \times 10^{-34}\;$&$\;2.29 \times 10^{-34}\;$\\

$10^{15}$&$\;6.86 \times 10^{-34}\;$&$\;7.62 \times 10^{-34}\;$&$\;8.22 \times 10^{-34}\;$&$\;8.82 \times 10^{-34}\;$&$\;9.28 \times 10^{-34}\;$\\

$10^{16}$&$\;1.95 \times 10^{-33}\;$&$\;2.19 \times 10^{-33}\;$&$\;2.45 \times 10^{-33}\;$&$\;2.79 \times 10^{-33}\;$&$\;3.12 \times 10^{-33}\;$\\

$10^{17}$&$\;4.93 \times 10^{-33}\;$&$\;5.54 \times 10^{-33}\;$&$\;6.33 \times 10^{-33}\;$&$\;7.49 \times 10^{-33}\;$&$\;8.75 \times 10^{-33}\;$\\

$10^{18}$&$\;1.14 \times 10^{-32}\;$&$\;1.28 \times 10^{-32}\;$&$\;1.48 \times 10^{-32}\;$&$\;1.79 \times 10^{-32}\;$&$\;2.16 \times 10^{-32}\;$\\

$10^{19}$&$\;2.44 \times 10^{-32}\;$&$\;2.73 \times 10^{-32}\;$&$\;3.19 \times 10^{-32}\;$&$\;3.95 \times 10^{-32}\;$&$\;4.87 \times 10^{-32}\;$\\

$10^{20}$&$\;4.76 \times 10^{-32}\;$&$\;5.27 \times 10^{-32}\;$&$\;6.22 \times 10^{-32}\;$&$\;7.89 \times 10^{-32}\;$&$\;9.99 \times 10^{-32}\;$\\
%

\hline
\end{tabular}
\caption{Total cross section as a function of $E_{\nu}$, with
$E_{\cal N}=M_{\cal N}$. We compare $\sigma^{tot}_{sm}$ with
$\sigma^{tot}_{2hdm}$ by taking: I(a) $M_{H^{\pm}}=90$ GeV and
$\tan\beta=60$; I(b) $M_{H^{\pm}}=150$ GeV and $\tan\beta=100$; I(c)
$M_{H^{\pm}}=225$ GeV and $\tan\beta=150$; I(d) $M_{H^{\pm}}=300$
GeV and $\tan\beta=200$.}\label{tab:t1}
\end{table}

\begin{table}
\begin{tabular}{|c|c|c|c|c|c|}
\hline
& & \multicolumn{4}{c|}{$\sigma^{tot}_{2hdm}$(cm$^2$)}\\
\cline{3-6} {$E_{\nu}$(eV)}& $\sigma^{tot}_{sm}$(cm$^2$) & II(a)
& II(b) & II(c) & II(d) \\
\hline \hline

$10^{14}$&$\;2.01 \times 10^{-34}\;$&$\;2.06 \times 10^{-34}\;$&$\;2.09 \times 10^{-34}\;$&$\;2.10 \times 10^{-34}\;$&$\;2.10 \times 10^{-34}\;$\\

$10^{15}$&$\;6.86 \times 10^{-34}\;$&$\;7.14 \times 10^{-34}\;$&$\;7.43 \times 10^{-34}\;$&$\;7.63 \times 10^{-34}\;$&$\;7.77 \times 10^{-34}\;$\\

$10^{16}$&$\;1.95 \times 10^{-33}\;$&$\;2.04 \times 10^{-33}\;$&$\;2.18 \times 10^{-33}\;$&$\;2.32 \times 10^{-33}\;$&$\;2.45 \times 10^{-33}\;$\\

$10^{17}$&$\;4.93 \times 10^{-33}\;$&$\;5.16 \times 10^{-33}\;$&$\;5.61 \times 10^{-33}\;$&$\;6.14 \times 10^{-33}\;$&$\;6.69 \times 10^{-33}\;$\\

$10^{18}$&$\;1.14 \times 10^{-32}\;$&$\;1.19 \times 10^{-32}\;$&$\;1.31 \times 10^{-32}\;$&$\;1.46 \times 10^{-32}\;$&$\;1.63 \times 10^{-32}\;$\\

$10^{19}$&$\;2.44 \times 10^{-32}\;$&$\;2.55 \times 10^{-32}\;$&$\;2.83 \times 10^{-32}\;$&$\;3.21 \times 10^{-32}\;$&$\;3.66 \times 10^{-32}\;$\\

$10^{20}$&$\;4.76 \times 10^{-32}\;$&$\;4.96 \times 10^{-32}\;$&$\;5.55 \times 10^{-32}\;$&$\;6.41 \times 10^{-32}\;$&$\;7.47 \times 10^{-32}\;$\\
%

\hline
\end{tabular}
\caption{Total cross section as a function of $E_{\nu}$, with
$E_{\cal N}=M_{\cal N}$. We compare $\sigma^{tot}_{sm}$ with
$\sigma^{tot}_{2hdm}$ by taking: II(a) $M_{H^{\pm}}=100$ GeV and
$\tan\beta=50$; II(b) $M_{H^{\pm}}=200$ GeV and $\tan\beta=100$;
II(c) $M_{H^{\pm}}=300$ GeV and $\tan\beta=150$; II(d)
$M_{H^{\pm}}=400$ GeV and $\tan\beta=200$.}\label{tab:t2}
\end{table}

\begin{table}
\begin{tabular}{|c|c|c|c|c|c|}
\hline
& & \multicolumn{4}{c|}{$\sigma^{tot}_{2hdm}$(cm$^2$)}\\
\cline{3-6} {$E_{\nu}$(eV)}& $\sigma^{tot}_{sm}$(cm$^2$) & III(a)
& III(b) & III(c) & III(d) \\
\hline \hline

$10^{14}$&$\;2.01 \times 10^{-34}\;$&$\;2.02 \times 10^{-34}\;$&$\;2.03 \times 10^{-34}\;$&$\;2.03 \times 10^{-34}\;$&$\;2.03 \times 10^{-34}\;$\\

$10^{15}$&$\;6.86 \times 10^{-34}\;$&$\;6.95 \times 10^{-34}\;$&$\;7.01 \times 10^{-34}\;$&$\;7.05 \times 10^{-34}\;$&$\;7.07 \times 10^{-34}\;$\\

$10^{16}$&$\;1.95 \times 10^{-33}\;$&$\;1.98 \times 10^{-33}\;$&$\;2.03 \times 10^{-33}\;$&$\;2.06 \times 10^{-33}\;$&$\;2.09 \times 10^{-33}\;$\\

$10^{17}$&$\;4.93 \times 10^{-33}\;$&$\;5.02 \times 10^{-33}\;$&$\;5.17 \times 10^{-33}\;$&$\;5.33 \times 10^{-33}\;$&$\;5.50 \times 10^{-33}\;$\\

$10^{18}$&$\;1.14 \times 10^{-32}\;$&$\;1.16 \times 10^{-32}\;$&$\;1.20 \times 10^{-32}\;$&$\;1.25 \times 10^{-32}\;$&$\;1.31 \times 10^{-32}\;$\\

$10^{19}$&$\;2.44 \times 10^{-32}\;$&$\;2.49 \times 10^{-32}\;$&$\;2.59 \times 10^{-32}\;$&$\;2.73 \times 10^{-32}\;$&$\;2.89 \times 10^{-32}\;$\\

$10^{20}$&$\;4.76 \times 10^{-32}\;$&$\;4.85 \times 10^{-32}\;$&$\;5.08 \times 10^{-32}\;$&$\;5.41 \times 10^{-32}\;$&$\;5.80 \times 10^{-32}\;$\\
%

\hline
\end{tabular}
\caption{Total cross section as a function of $E_{\nu}$, with
$E_{\cal N}=M_{\cal N}$. We compare $\sigma^{tot}_{sm}$ with
$\sigma^{tot}_{2hdm}$ by taking: III(a) $M_{H^{\pm}}=150$ GeV and
$\tan\beta=50$; III(b) $M_{H^{\pm}}=300$ GeV and $\tan\beta=100$;
III(c) $M_{H^{\pm}}=450$ GeV and $\tan\beta=150$; III(d)
$M_{H^{\pm}}=600$ GeV and $\tan\beta=200$.}\label{tab:t3}
\end{table}

\begin{table}
\begin{tabular}{|c|c|c|c|c|}
\hline
 & \multicolumn{4}{c|}{$\sigma^{tot}_{H^+}/\sigma^{tot}_{sm}$}\\
\cline{2-5}
{$E_{\nu}$(eV)} & I(a) & I(b) & I(c) & I(d) \\

\hline \hline

$10^{14}\;$&$\;0.079\;$&$\;0.111\;$&$\;0.131\;$&$\;0.141\;$\\

$10^{15}\;$&$\;0.111\;$&$\;0.198\;$&$\;0.286\;$&$\;0.353\;$\\

$10^{16}\;$&$\;0.123\;$&$\;0.256\;$&$\;0.429\;$&$\;0.596\;$\\

$10^{17}\;$&$\;0.125\;$&$\;0.284\;$&$\;0.519\;$&$\;0.775\;$\\

$10^{18}\;$&$\;0.123\;$&$\;0.298\;$&$\;0.576\;$&$\;0.899\;$\\

$10^{19}\;$&$\;0.119\;$&$\;0.306\;$&$\;0.617\;$&$\;0.995\;$\\

$10^{20}\;$&$\;0.109\;$&$\;0.308\;$&$\;0.659\;$&$\;1.100\;$\\
%

\hline
\end{tabular}
\caption{$\sigma^{tot}_{H^+}(\nu_{\tau} + {\cal N} \to \tau^{-} +
X)/\sigma^{tot}_{sm}(\nu_{\tau} + {\cal N} \to \tau^{-} + X)$ as a
function of $E_{\nu}$, with $E_{\cal N}=M_{\cal N}$ for the cases:
I(a) $M_{H^{\pm}}=90$ GeV and $\tan\beta=60$; I(b) $M_{H^{\pm}}=150$
GeV and $\tan\beta=100$; I(c) $M_{H^{\pm}}=225$ GeV and
$\tan\beta=150$; I(d) $M_{H^{\pm}}=300$ GeV and
$\tan\beta=200$.}\label{tab:t4}
\end{table}

\begin{table}
\begin{tabular}{|c|c|c|c|c|}
\hline
 & \multicolumn{4}{c|}{$\sigma^{tot}_{H^+}/\sigma^{tot}_{sm}$}\\
\cline{2-5}
{$E_{\nu}$(eV)} & II(a) & II(b) & II(c) & II(d) \\

\hline \hline

$10^{14}\;$&$\;0.027\;$&$\;0.040\;$&$\;0.045\;$&$\;0.047\;$\\

$10^{15}\;$&$\;0.040\;$&$\;0.082\;$&$\;0.112\;$&$\;0.132\;$\\

$10^{16}\;$&$\;0.046\;$&$\;0.118\;$&$\;0.189\;$&$\;0.254\;$\\

$10^{17}\;$&$\;0.047\;$&$\;0.139\;$&$\;0.245\;$&$\;0.358\;$\\

$10^{18}\;$&$\;0.047\;$&$\;0.151\;$&$\;0.284\;$&$\;0.436\;$\\

$10^{19}\;$&$\;0.046\;$&$\;0.160\;$&$\;0.315\;$&$\;0.499\;$\\

$10^{20}\;$&$\;0.043\;$&$\;0.168\;$&$\;0.348\;$&$\;0.571\;$\\
%

\hline
\end{tabular}
\caption{$\sigma^{tot}_{H^+}(\nu_{\tau} + {\cal N} \to \tau^{-} +
X)/\sigma^{tot}_{sm}(\nu_{\tau} + {\cal N} \to \tau^{-} + X)$ as a
function of $E_{\nu}$, with $E_{\cal N}=M_{\cal N}$ for the cases:
II(a) $M_{H^{\pm}}=100$ GeV and $\tan\beta=50$; II(b)
$M_{H^{\pm}}=200$ GeV and $\tan\beta=100$; II(c) $M_{H^{\pm}}=300$
GeV and $\tan\beta=150$; II(d) $M_{H^{\pm}}=400$ GeV and
$\tan\beta=200$.}\label{tab:t5}
\end{table}

\begin{table}
\begin{tabular}{|c|c|c|c|c|}
\hline
 & \multicolumn{4}{c|}{$\sigma^{tot}_{H^+}/\sigma^{tot}_{sm}$}\\
\cline{2-5}
{$E_{\nu}$(eV)} & III(a) & III(b) & III(c) & III(d) \\

\hline \hline

$10^{14}\;$&$\;0.007\;$&$\;0.009\;$&$\;0.009\;$&$\;0.010\;$\\

$10^{15}\;$&$\;0.012\;$&$\;0.022\;$&$\;0.028\;$&$\;0.031\;$\\

$10^{16}\;$&$\;0.016\;$&$\;0.037\;$&$\;0.056\;$&$\;0.072\;$\\

$10^{17}\;$&$\;0.018\;$&$\;0.048\;$&$\;0.082\;$&$\;0.116\;$\\

$10^{18}\;$&$\;0.019\;$&$\;0.056\;$&$\;0.102\;$&$\;0.152\;$\\

$10^{19}\;$&$\;0.019\;$&$\;0.062\;$&$\;0.118\;$&$\;0.184\;$\\

$10^{20}\;$&$\;0.019\;$&$\;0.069\;$&$\;0.137\;$&$\;0.220\;$\\
%

\hline
\end{tabular}
\caption{$\sigma^{tot}_{H^+}(\nu_{\tau} + {\cal N} \to \tau^{-} +
X)/\sigma^{tot}_{sm}(\nu_{\tau} + {\cal N} \to \tau^{-} + X)$ as a
function of $E_{\nu}$, with $E_{\cal N}=M_{\cal N}$ for the cases:
III(a) $M_{H^{\pm}}=150$ GeV and $\tan\beta=50$; III(b)
$M_{H^{\pm}}=300$ GeV and $\tan\beta=100$; III(c) $M_{H^{\pm}}=450$
GeV and $\tan\beta=150$; III(d) $M_{H^{\pm}}=600$ GeV and
$\tan\beta=200$.}\label{tab:t6}
\end{table}

\begin{table}
\begin{tabular}{|c|c|c|c|c|c|c|}
\hline
& \multicolumn{6}{c|}{$\sigma^{tot}_{H^+}$(cm$^2$)}\\
\cline{2-7} {$E_{\nu}$(eV)}& all quarks & bottom & strange
& anti-charm & down & anti-up\\
\hline \hline

$10^{14}$&$\;9.40 \times 10^{-36}\;$&$\;9.37 \times 10^{-36}\;$&$\;1.67 \times 10^{-38}\;$&$\;1.20 \times 10^{-38}\;$&$\;3.15 \times 10^{-40}\;$&$\;8.97 \times 10^{-41}\;$\\

$10^{15}$&$\;9.06 \times 10^{-35}\;$&$\;9.03 \times 10^{-35}\;$&$\;1.25 \times 10^{-37}\;$&$\;9.97 \times 10^{-38}\;$&$\;1.35 \times 10^{-39}\;$&$\;6.12 \times 10^{-40}\;$\\

$10^{16}$&$\;4.96 \times 10^{-34}\;$&$\;4.95 \times 10^{-34}\;$&$\;5.45 \times 10^{-37}\;$&$\;4.86 \times 10^{-37}\;$&$\;3.18 \times 10^{-39}\;$&$\;2.40 \times 10^{-39}\;$\\

$10^{17}$&$\;1.76 \times 10^{-33}\;$&$\;1.76 \times 10^{-33}\;$&$\;1.67 \times 10^{-36}\;$&$\;1.57 \times 10^{-36}\;$&$\;7.26 \times 10^{-39}\;$&$\;6.86 \times 10^{-39}\;$\\

$10^{18}$&$\;4.96 \times 10^{-33}\;$&$\;4.95 \times 10^{-33}\;$&$\;4.27 \times 10^{-36}\;$&$\;4.10 \times 10^{-36}\;$&$\;1.73 \times 10^{-38}\;$&$\;1.72 \times 10^{-38}\;$\\

$10^{19}$&$\;1.22 \times 10^{-32}\;$&$\;1.22 \times 10^{-32}\;$&$\;9.93 \times 10^{-36}\;$&$\;9.60 \times 10^{-36}\;$&$\;3.96 \times 10^{-38}\;$&$\;3.95 \times 10^{-38}\;$\\

$10^{20}$&$\;2.72 \times 10^{-32}\;$&$\;2.71 \times 10^{-32}\;$&$\;2.14 \times 10^{-35}\;$&$\;2.07 \times 10^{-35}\;$&$\;8.45 \times 10^{-38}\;$&$\;8.45 \times 10^{-38}\;$\\
%

\hline
\end{tabular}
\caption{Contribution to $\sigma^{tot}_{H^+}$ from the different
allowed initial quarks as a function of $E_{\nu}$, with $E_{\cal
N}=M_{\cal N}$, taking $M_{H^{\pm}}=400$ GeV and $\tan\beta=200$.}
\label{tab:t7}
\end{table}

\begin{table}
\begin{tabular}{|c|c|c|c|c|c|}
\hline
& & \multicolumn{4}{c|}{$\langle y\rangle^{CC}_{2hdm}$}\\
\cline{3-6} {$E_{\nu}$(eV)}& $\;\langle y\rangle^{CC}_{sm}\;$ & I(a)
& I(b) & I(c) & I(d) \\
\hline \hline

$10^{14}$&$\;0.355\;$&$\;0.383\;$&$\;0.394\;$&$\;0.401\;$&$\;0.405\;$\\

$10^{15}$&$\;0.289\;$&$\;0.332\;$&$\;0.362\;$&$\;0.388\;$&$\;0.407\;$\\

$10^{16}$&$\;0.255\;$&$\;0.305\;$&$\;0.349\;$&$\;0.394\;$&$\;0.430\;$\\

$10^{17}$&$\;0.234\;$&$\;0.286\;$&$\;0.339\;$&$\;0.397\;$&$\;0.443\;$\\

$10^{18}$&$\;0.218\;$&$\;0.272\;$&$\;0.330\;$&$\;0.396\;$&$\;0.449\;$\\

$10^{19}$&$\;0.204\;$&$\;0.258\;$&$\;0.321\;$&$\;0.394\;$&$\;0.453\;$\\

$10^{20}$&$\;0.180\;$&$\;0.232\;$&$\;0.302\;$&$\;0.386\;$&$\;0.452\;$\\
%

\hline
\end{tabular}
\caption{Mean inelasticity parameter for charged current interaction
$\langle y \rangle^{CC}$ as a function of $E_{\nu}$, with $E_{\cal
N}=M_{\cal N}$. We compare $\langle y \rangle^{CC}_{sm}$ with
$\langle y \rangle^{CC}_{2hdm}$ by taking: I(a) $M_{H^{\pm}}=90$ GeV
and $\tan\beta=60$; I(b) $M_{H^{\pm}}=150$ GeV and $\tan\beta=100$;
I(c) $M_{H^{\pm}}=225$ GeV and $\tan\beta=150$; I(d)
$M_{H^{\pm}}=300$ GeV and $\tan\beta=200$.}\label{tab:t8}
\end{table}


\begin{references}

\bibitem{stanmod} S. L. Glashow, Nucl. Phys. {\bf 22}, 579 (1961); S.
Weinberg, Phys. Rev. Lett. {\bf 19}, 1264 (1967); A. Salam, Proc.
8th NOBEL Symposium, ed. N. Svartholm (Almqvist and Wiksell,
Stockholm, 1968), p. 367.

\bibitem{HHG}
J. Gunion, H. Haber, G. Kane and S. Dawson, {\it The Higgs Hunter's
Guide}, Addison-Wesley Publishig Company, Reading, MA, 1990.

\bibitem{TSM}
H. E. Haber, in {\it Testing the Standard Model}, Proceedings of the
1990 Theoretical Advanced Study Institute in Elementary Particle
Physics, edited by M. Cvetic and P. Langacker (World Scientific,
Singapore, 1991) p. 340-475.

\bibitem{Barger:1989fj}
  V.~D.~Barger, J.~L.~Hewett and R.~J.~N.~Phillips,
  Phys.\ Rev.\ D {\bf 41}, 3421 (1990).

\bibitem{cpv}
  T.~D.~Lee,
  Phys.\ Rev.\ D {\bf 8}, 1226 (1973);
  T.~D.~Lee,
  Phys.\ Rept.\  {\bf 9}, 143 (1974);
  S.~Weinberg,
  Phys.\ Rev.\ Lett.\  {\bf 37}, 657 (1976);
  Y.~L.~Wu and L.~Wolfenstein,
  Phys.\ Rev.\ Lett.\  {\bf 73}, 1762 (1994)
  [arXiv:hep-ph/9409421];
  J.~Liu and L.~Wolfenstein,
  Nucl.\ Phys.\ B {\bf 289}, 1 (1987).

\bibitem{Heister:2002ev}
  A.~Heister {\it et al.}  [ALEPH Collaboration],
  Phys.\ Lett.\ B {\bf 543}, 1 (2002)
  [arXiv:hep-ex/0207054].


\bibitem{Ackerstaff:1998yk}
  K.~Ackerstaff {\it et al.}  [OPAL Collaboration],
  Eur.\ Phys.\ J.\ C {\bf 8}, 3 (1999)
  [arXiv:hep-ex/9808016].

\bibitem{Abbiendi:2002jw}
  G.~Abbiendi {\it et al.}  [OPAL Collaboration],
  Phys.\ Lett.\ B {\bf 551}, 35 (2003)
  [arXiv:hep-ex/0211066].

\bibitem{Abbiendi:2001fi}
  G.~Abbiendi {\it et al.}  [OPAL Collaboration],
  Phys.\ Lett.\ B {\bf 520}, 1 (2001)
  [arXiv:hep-ex/0108031].

\bibitem{Ammar:1996xh}
  R.~Ammar {\it et al.}  [CLEO Collaboration],
  Phys.\ Rev.\ Lett.\  {\bf 78}, 4686 (1997).

\bibitem{Stahl:1996gu}
  A.~Stahl and H.~Voss,
  Z.\ Phys.\ C {\bf 74}, 73 (1997).

\bibitem{Buskulic:1994gj}
  D.~Buskulic {\it et al.}  [ALEPH Collaboration],
  Phys.\ Lett.\ B {\bf 343}, 444 (1995).

\bibitem{Barate:2000rc}
  R.~Barate {\it et al.}  [ALEPH Collaboration],
  Eur.\ Phys.\ J.\ C {\bf 19}, 213 (2001)
  [arXiv:hep-ex/0010022].

\bibitem{Acciarri:1996bv}
  M.~Acciarri {\it et al.}  [L3 Collaboration],
  Phys.\ Lett.\ B {\bf 396}, 327 (1997).

\bibitem{Roy:2005yu}
  D.~P.~Roy,
  AIP Conf.\ Proc.\  {\bf 805}, 110 (2006)
  [arXiv:hep-ph/0510070].

\bibitem{Baer:2002hf}
  H.~Baer, J.~Ferrandis and X.~Tata,
  Phys.\ Lett.\ B {\bf 561}, 145 (2003)
  [arXiv:hep-ph/0211418].

\bibitem{telescopes} {\it AMANDA Collaboration}, E. Andres {\it et al.},
Nature {\bf 410}, 441 (2001); {\it ANTARES Collaboration}, Y.
Becherini {\it et al.}, e-Print Archive: hep-ph/0211173; {\it AUGER
Collaboration}, D. Zavrtanik {\it et al.}, Nucl. Phys. Proc. Suppl.
{\bf 85}, 324 (2002); {\it NESTOR Collaboration}, P. K. F. Grieder
{\it et al.}, Nuovo Cim. {\bf 24C}, 771 (2001); {\it RICE
Collaboration}, I. Kravchenko {\it et al.}, Astropart. Phys. {\bf
19}, 15 (2003).

\bibitem{uhe-neut}
  V.~S.~Beresinsky and G.~T.~Zatsepin,
  Phys.\ Lett.\ B {\bf 28}, 423 (1969);
  V.~S.~Berezinsky and V.~I.~Dokuchaev,
  Nucl.\ Phys.\ Proc.\ Suppl.\  {\bf 110}, 522 (2002);
  V.~S.~Berezinsky,
  Nucl.\ Phys.\ Proc.\ Suppl.\  {\bf 38}, 363 (1995);
 and
  Nucl.\ Phys.\ Proc.\ Suppl.\  {\bf 31}, 413 (1993);
  T.~Stanev,
  Nucl.\ Phys.\ Proc.\ Suppl.\  {\bf 14A}, 17 (1990);
  K.~Greisen,
  Phys.\ Rev.\ Lett.\  {\bf 16}, 748 (1966);
  C.~T.~Hill and D.~N.~Schramm,
  Phys.\ Lett.\ B {\bf 131}, 247 (1983);
  and
  Phys.\ Rev.\ D {\bf 31}, 564 (1985).

\bibitem{Gandhi:1998ri}
  R.~Gandhi, C.~Quigg, M.~H.~Reno and I.~Sarcevic,
  Phys.\ Rev.\ D {\bf 58}, 093009 (1998)
  [arXiv:hep-ph/9807264].

\bibitem{Blanch1}
O.~Blanch and P.~Billoir, "Acceptance and Flux Limit for
$\nu_{\tau}$ with the Pierre Auger Observatory Surface Detector",
Preprint LPNHE, Paris (France), September 26, 2005.

\bibitem{Bertou:2001vm}
  X.~Bertou, P.~Billoir, O.~Deligny, C.~Lachaud and A.~Letessier-Selvon,
  Astropart.\ Phys.\  {\bf 17}, 183 (2002)
  [arXiv:astro-ph/0104452].

\bibitem{partmod1} R. P. Feynman: Photon-hadron interactions.
Reading: Benjamin 1972.

\bibitem{partmod2} V.D. Barger and R.J.N. Phillips, Collider Physics (Updated Edition),
Addison-Wesley Publishing Company, Inc., Reading, Massachusetts,
1997.

\bibitem{pumplin} J. Pumplin {\it et al.}, JHEP {\bf 207}, 12 (2002);
D. Stump {\it et al.}, e-Print Archive: hep-ph/0303013.

\bibitem{Kinematics}E. Byckling and Kajantie: Particle kinematics.
New York: Willey 1972.

\bibitem{dis-sm}
  M.~H.~Reno and C.~Quigg,
  Phys.\ Rev.\ D {\bf 37}, 657 (1988);
  C.~Quigg, M.~H.~Reno and T.~P.~Walker,
  Phys.\ Rev.\ Lett.\  {\bf 57}, 774 (1986).

\bibitem{Rosado:2006hn}
  A.~Rosado,
  Phys.\ Rev.\ D {\bf 74}, 057301 (2006).

\bibitem{partdata}
S.~Eidelman {\it et al.}  [Particle Data Group],
Phys.\ Lett.\ B {\bf 592}, 1 (2004).

\bibitem{Palomares-Ruiz:2005xw}
  S.~Palomares-Ruiz, A.~Irimia and T.~J.~Weiler,
  Phys.\ Rev.\ D {\bf 73}, 083003 (2006)
  [arXiv:astro-ph/0512231].

\bibitem{HIRES}
  P.~Sokolsky and J.~Belz  [the HiRes Collaboration],
  ``Comparison of UHE composition measurements by Fly's Eye, HiRes-prototype /
  MIA and stereo HiRes experiments,''
  arXiv:astro-ph/0507485.

\bibitem{Augerfluo}
  H.~Blumer {\it et al.}  [Auger Collaboration],
  ``The Auger fluorescence detector prototype telescope,''
FZKA-6345P
{\it Prepared for 26th International Cosmic Ray Conference (ICRC
99), Salt Lake City, UT, 17-25 Aug 1999};
  J.~Abraham {\it et al.}  [Pierre Auger Collaboration],
  Nucl.\ Instrum.\ Meth.\ A {\bf 523}, 50 (2004).

\bibitem{TAfluo}
  Z.~Cao,
  Nucl.\ Phys.\ Proc.\ Suppl.\  {\bf 151}, 287 (2006).

\bibitem{EUSO}
  Ph.~Gorodetzky  [The EUSO Collaboration],
  Nucl.\ Phys.\ Proc.\ Suppl.\  {\bf 151}, 401 (2006)
  [arXiv:astro-ph/0502187];
  G.~D'Ali Staiti  [EUSO Collaboration],
  Nucl.\ Phys.\ Proc.\ Suppl.\  {\bf 136}, 415 (2004).

\bibitem{OWL}
  F.~W.~Stecker, J.~F.~Krizmanic, L.~M.~Barbier, E.~Loh, J.~W.~Mitchell, P.~Sokolsky and R.~E.~Streitmatter,
  Nucl.\ Phys.\ Proc.\ Suppl.\  {\bf 136C}, 433 (2004)
  [arXiv:astro-ph/0408162].

\bibitem{Zas:2005zz}
  E.~Zas,
  New J.\ Phys.\  {\bf 7}, 130 (2005)
  [arXiv:astro-ph/0504610].

\bibitem{Gandhi:1995tf}
  R.~Gandhi, C.~Quigg, M.~H.~Reno and I.~Sarcevic,
  Astropart.\ Phys.\  {\bf 5}, 81 (1996)
  [arXiv:hep-ph/9512364].

\bibitem{Arteaga:1}
  J.~C.~Arteaga-Velazquez and A.~Zepeda,
   ``Estimation of the detectable flux of astrophysical neutrinos at the Pierre
   Auger observatory by means of horizontal air showers,''
{\it Proceedings of 29th International Cosmic Ray Conference Pune
{\bf Vol. 9}, 151-154 (2005).}

\bibitem{Aramo:2004pr}
  C.~Aramo, A.~Insolia, A.~Leonardi, G.~Miele, L.~Perrone, O.~Pisanti and D.~V.~Semikoz,
  Astropart.\ Phys.\  {\bf 23}, 65 (2005)
  [arXiv:astro-ph/0407638].

\end{references}
\end{document}